\documentstyle[amssymb,aps]{revtex}

\begin{document}
\draft
\title{Order $N$ photonic band structures for metals and other dispersive
materials}
\author{J. Arriaga}
\address{Instituto de F\'{\i}sica, Universidad Aut\'onoma
de Puebla \\ Apartado Postal J-48, 72570, Puebla, M\'exico}
\author{A.J. Ward and J.B. Pendry}
\address{The Blackett Laboratory, Imperial College
 London SW7 2BZ, United Kingdom}
\date{\today}
\maketitle

\begin{abstract}
We show, for the first time, how to calculate photonic band structures
for metals and other
dispersive systems using an efficient Order $N$ scheme. The method is
applied to two simple periodic metallic systems where it gives
results in close 
agreement with calculations made with other techniques. Further, the
approach demonstrates excellent numerical stablity within the limits
we give.
Our new method opens the way for efficient calculations on complex
structures containing a whole new class of material.
\end{abstract}

\pacs{PACS numbers: 42.70.Qs, 02.70.Bf, 78.20.Bh}

In recent years, there has been much interest in artificially structured
dielectrics, otherwise known as photonic
crystals~\cite{ASI}. Structured on the scale of the wavelength of light, these
materials promise new and exciting optical properties based on the novel
dispersion relationships, $\omega(k)$, induced by the periodic
structure. There is a strong analogy here with semiconductor physics,
where
band gaps in the dispersion relationships for electrons play a key role
in determining the electronic properties. This analogy has been emphasized
by many authors~\cite{Yab,JcondMat}.

Conceptually, the problem reduces to solving Maxwell's equations,

\begin{equation}
\nabla\times{\bf E}=-\frac{\partial{\bf B}}{\partial t}\hspace{1cm}
\nabla\times{\bf H}=+\frac{\partial{\bf D}}{\partial t},
\end{equation}
where the photonic structure may be introduced either through the electric
permittivity, ${\bf D}({\bf r},\omega)=\varepsilon({\bf r},\omega)
{\bf E}({\bf r},\omega)$, or, less commonly,
the magnetic permeability, ${\bf H}({\bf r},\omega)
=\mu({\bf r},\omega){\bf B}({\bf r},\omega)$. In reciprocal space we have, 
\begin{equation}
i{\bf k}\times {\bf E}({\bf k},\omega )=+i\omega {\bf B}({\bf k},\omega
),\;\;\;i{\bf k}\times {\bf H}({\bf k},\omega )=-i\omega {\bf D}({\bf k}
,\omega ).
\label{eq:maxk}
\end{equation}
These equations provide the basis for several computational schemes. 
Theorists have deployed three main techniques for calculation of $\omega
\left( k\right) $: plane wave expansions~\cite{pw1,pw2,pw3}, 
transfer matrix methods in real space~\cite{tmm1,tmm2}, 
and, more recently, the so called `Order~$N$' method of 
Chan, Yu and Ho~\cite{OrderN}. 
Currently, the field is dominated by the latter two methods, each
having its own strengths. 
As in the electronic case computation times
for traditional schemes, such as plane wave expansions, scale as $N^3$, 
where $N$ is proportional to the
size of the system. The scaling law follows from the O($N^3$) operations
required to diagonalize an $N\times N$ matrix. This unfavorable scaling with
system size has proved very restrictive and there has been an increase in
activity to devise new methods which scale more favorably with the system size.
The optimum scaling possible is clearly O($N$) -- the time taken to define
the problem or to look at the answer! Chan {\em etal.} showed 
it is possible to
realize this optimal scaling by working in real space and in the time domain,
provided that the resulting equations are local in space and time. That is
to say, the fields at each point in space-time can be updated using only the
recent values of the fields at points which are close by in real space. In
the context of photonic band structure calculations, this amounts to
requiring that the dielectric function and the magnetic permeability are
constants independent of $k$ and $\omega$. 
In fact, Chan's method was not a new idea but had been known for some time
to the electrical engineering community as the `finite difference time domain
(FDTD)' method~\cite{FDTD}, though it had not previously been applied to
the photonic band structure problem.

However, there is a problem with methods such as FDTD.
In the time domain, it is not obvious how to deal with a frequency
dependent dielectric permittivity, $\varepsilon(\omega)$.
For example, nearly-free-electron metals can be modeled by,
\begin{equation}
\varepsilon (\omega )=1-\frac{\omega _p^2}{\omega ^2},
\label{eq:plasma}
\end{equation}
where $\omega _p$ is the plasma frequency. 
This restriction is a major obstacle to progress in the field as
Order~$N$ methods are ideal for treating complex unit cells but are excluded
from attacking some of the most interesting systems. Metals, for example, 
show some of the most striking effects when micro-structured. 
Earlier calculations using the
transfer matrix method show huge enhancements of local fields in
nanostructured arrays of metal cylinders or spheres~\cite{sers}.
But despite the clear advantage that an O(N) scaling would give in this
calculation, it is impossible to do using a standard Order~$N$ scheme.

The electrical engineering community have known about this problem for
some time and progress has been made towards solving it~\cite{Lueb2,Taf}.
Here we present a method similar to that of Luebbers~\cite{Lueb} to
calculate for the first time the band structure of various photonic 
structures containing metallic components in an Order~$N$ scheme.
Specifically, we treat
metallic systems exhibiting a simple plasmon pole, 
as given in  equation~(\ref{eq:plasma}). Importantly, this
can be done without destroying the O($N$) scaling property, and we argue
that our methods can be extended to treat more complex forms of dispersion.
This constitutes a substantial advance to Chan's original Order~$N$ band
structure scheme and allows us to tackle a whole new class of
materials previously excluded from the Order~$N$ method.

We begin with Maxwell's equations from~(\ref{eq:maxk}) and approximate 
them as follows,
\begin{equation}
i{\bf \kappa }_E\left( {\bf k}\right) \times {\bf E}({\bf k},\omega
)=+iw_H\left( \omega \right) \mu \mu _0{\bf H}({\bf k},\omega ),\;\;\;i{\bf
\kappa }_H\left( {\bf k}\right) \times {\bf H}({\bf k},\omega )=-iw_E\left(
\omega \right) \varepsilon \varepsilon _0{\bf E}({\bf k},\omega ),
\label{eq:max}
\end{equation}
where,
\begin{equation}
\begin{array}{ll}
{\bf \kappa }_{Ex}\left( k_x\right) =\left( +ia_0\right) ^{-1}\left[ \exp
\left( +ik_xa_0\right) -1\right] , & {\bf \kappa }_{Hx}\left( k_x\right)
=\left( -ia_0\right) ^{-1}\left[ \exp \left( -ik_xa_0\right) -1\right]
,\quad \text{etc.} \\ 
w_E\left( \omega \right) =\left( -iT\right) ^{-1}\left[ \exp \left( -i\omega
T\right) -1\right] , & w_H\left( \omega \right) =\left( +iT\right)
^{-1}\left[ \exp \left( +i\omega T\right) -1\right].
\end{array}
\end{equation}
In the case that $\varepsilon$ and $\mu$ are independent of $\omega$,
Fourier transformation to $({\bf r},t)$ space yields a set of
finite difference equations on a simple cubic mesh in real space, of lattice
constant $a_0$, and on an interval $T$ in the time domain, as discussed 
in~\cite{JcondMat,Green}. 
The resulting equations allow us to update the ${\bf H}$ and ${\bf E}$
fields in the time domain. Starting from, say, a random set of fields and
the appropriate periodic boundary conditions we iterate for a sufficient
number of
time steps, determined by the frequency resolution desired, and then perform a
Fourier transform to obtain the frequencies, $\omega({\bf k}) $.

In this paper, we wish to consider band structures for systems in which the 
permittivity is a not a constant but a function of frequency,

\begin{equation}
i{\bf \kappa }_E\left( {\bf k}\right) \times {\bf E}({\bf k},\omega
)=+iw_H\left( \omega \right) \mu _0{\bf H}({\bf k},\omega ),\;\;\;i{\bf %
\kappa }_H\left( {\bf k}\right) \times {\bf H}({\bf k},\omega )=-iw_E\left(
\omega \right) \varepsilon \left( \omega \right) \varepsilon _0{\bf E}({\bf k%
},\omega ).
\end{equation}
The second equation can be written as 
\begin{equation}
\varepsilon {^{-1}}\left( {\omega }\right) i{\bf \kappa }_H\left( {\bf k}
\right) \times {\bf H}\left( {\bf k},\omega \right) =-iw_E\left( \omega
\right) \varepsilon _0{\bf E}\left( {\bf k},\omega \right). 
\end{equation}
We substitute the single pole metallic form from~(\ref{eq:plasma}) and 
approximate,
\begin{eqnarray}
\varepsilon^{-1}(\omega) -1 &=&\frac{\omega _p^2}{\omega
^2-\omega _p^2}  \nonumber \\
&\thickapprox &\frac{1}{2}
\left[ 
 \frac{-iT\omega _p}{1-\exp \left[ +i\left(\omega-\omega_p\right) T\right]}
-\frac{-iT\omega _p}{1-\exp \left[ +i\left(\omega+\omega_p\right) T\right]}
\right]\nonumber \\
&=&\frac{T\omega _p}{2i}\left[
\sum_{n=0}^\infty \exp \left[ i\left( \omega
-\omega _p\right) nT\right]
-\sum_{n=0}^\infty \exp \left[ i\left( \omega
+\omega _p\right) nT\right] \right],
\label{eq:approx}
\end{eqnarray}
and Fourier transform to give, 
\begin{eqnarray}
\varepsilon_0{\bf E}\left({\bf k},t+T\right)&=&\varepsilon_0{\bf E}
\left({\bf k},t\right)+iT{\bf \kappa }_H\left( {\bf k}\right) \times {\bf H}
\left({\bf k},t\right) \nonumber \\
&&-\frac{T^2\omega_p}{2}\left[U_{+}(t)-U_{-}(t)\right],
\label{eq:update1}
\end{eqnarray}
where, 
\begin{equation}
U_{\pm }(t+T)={\bf \kappa }_H\left( {\bf k}\right) \times {\bf H}\left( {\bf
r},t+T\right) +\exp \left( \pm i\omega _pT\right) U_{\pm }(t).
\label{eq:FD2TD}
\end{equation}
When transformed into real space, these equations together with 
\begin{equation}
\mu_0{\bf H}\left( {\bf k},t+T\right) =\mu_0{\bf H}\left( {\bf k},t\right)
-iT{\bf \kappa }_E\left( {\bf k}\right) \times {\bf E}\left( {\bf r}
,t\right) ,
\label{eq:update2}
\end{equation}
give us the expressions for the frequency dependent Order~$N$ method.
The third term of equation~(\ref{eq:FD2TD})
gives the contribution due to the frequency-dependent dielectric
constant. It should be noted that this term includes a non-local time effect,
and is a function of the previous history of the system.

By expanding equation~(\ref{eq:approx}) in powers of $\omega_{p}T$ it can be
shown that the approximation we have made is good for frequencies above about
 $\omega_{p}^{2}T/\sqrt{12}$. In general, the Order~$N$ method 
is valid up to frequencies of about $c_{0}\pi/a_{0}$, with the usual
stability critirion that $T\leq a_{0}/\sqrt{3}c_{0}$. 
Together, these define the limits of applicability of our approach.
In fact, an expansion of equation~(\ref{eq:approx}) can be used to
produce systematic corrections to the approximation and obtain an
arbitrary lower frequency limit. In our calculations we found that it
was sufficient just to correct for the leading error.

In general, a dielectric function that depends on $\omega $ would give
equations
which require us to remember all previous history of the calculation in
order to update the fields, and all computational advantage is lost for the
scheme. However, for the special case where $\varepsilon^{-1}(\omega)$ 
can be represented by a simple pole, as in equation~(\ref{eq:approx}),
the fields may be rapidly updated by defining $U_{\pm }(t)$ which themselves
may be updated from a knowledge of the immediate past only, as shown 
in~(\ref{eq:FD2TD}). 
This stems from the form of the approximation to the pole which we chose 
in~(\ref{eq:approx}).
Clearly, this result can easily be generalized to the case where 
$\varepsilon^{-1}(\omega)$ can be represented by a sum over poles. The poles
have a simple interpretation of internal electromagnetically active modes
which can absorb energy. Many inverse dielectric functions are well
approximated by a sum over poles, particularly where the poles lie off the
real axis and therefore give rise to rather broad structure in 
 $\varepsilon^{-1}(\omega)$. In so far as dielectric 
functions can be
approximated in this manner, we have a general technique for including
dispersion in an Order~$N$ calculation.

We now present results for the photonic band structure for both
one and two dimensional systems. In
Fig.~\ref{fig:1d}, we show the photonic
band structure for a one-dimensional system formed of alternating layers of
a metal of thickness $a$ separated by a vacuum layers of thickness 
$b$, the unit cell for which we divide up into ten equally spaced mesh
points.
The metal was characterized by a single pole plasma as given by
equation~(\ref{eq:approx}), and we considered propagation  at normal
incidence. We chose $\omega _pd/2\pi c_0=1,\quad d=a+b,\quad $,
 $f=0.1$, where $w_p$ is the
plasma frequency and $f$ is the metal filling fraction. A total of $8192$
time steps were used in the simulation, with each time step of $0.64$ in 
units of $m_{e}a_{0}^{2}/\hbar$.
From Fig.~\ref{fig:1d} we can see that allowed
bands exist in the frequency range $\omega/\omega_{p}\leq 1$ in which the
dielectric function of the metallic component is negative. The most
interesting feature of the band structure is the gap below the lowest
frequency band which exists for any filling fraction. Also shown in 
Fig.~\ref{fig:1d_tmm} are results for the same system obtained with the
transfer matrix method for comparison. The excellent agreement gives
confidence that the modified Order~$N$ method is working correctly.

In Fig.~\ref{fig:gap} we plot
the width of the gap below the lowest frequency band as a function of the
filling fraction. From this figure, we see that the width of this gap
increases with the filling fraction. This result agrees with
previously published results~\cite{ASI,maradudin}, again confirming the
validity of our method. 

For the two dimensional case we calculate the photonic band structures for
a system consisting of an infinite array of parallel, metal
rods of square cross-section,  embedded in vacuum. The rods are
infinitely  long in the $z$-direction and arranged on a square lattice
in the $x-y$ plane.
The thickness
of the rods is $a$, and their separation $d$, and our unit cell now 
consists of
$10\times 10$ mesh points. For the ${\bf E}$ and ${\bf H}$
-polarizations we have ${\bf E}\left( {\bf r},t\right) =\left[ 
\begin{array}{lll}
0, & 0, & E_3\left( {\bf r},t\right)
\end{array}
\right] $ and ${\bf H}\left( {\bf r},t\right) =\left[ 
\begin{array}{lll}
0, & 0, & H_3\left( {\bf r},t\right)
\end{array}
\right] $ respectively.
In Fig.~\ref{fig:2d}, we present the photonic band
structure  when the filling fraction $f=0.01$. Figure~\ref{fig:2dH}
show the result for the photonic band structure
when the filling fraction of the rods is $f=0.04$. In both cases we
again show transfer matrix results for comparison, which again confirm
the accuracy of the new method.
The key features of these results are the low frequency cut-off in the 
 ${\bf E}$-polarized bands and the very flat bands at around the
plasma frequency in the ${\bf H}$-polarization. The cut-off is caused by the
collective motion of electrons screening the electric field parallel to the 
rods, below some effective plasma frequency determined
by the filling fraction. The flat bands are caused by the resonant modes 
of the individual rods, excited by the electric field perpendicular to the
rods.
The reproduction of these previously observed effects again confirms that
our approach is capturing the correct physics.

We have calculated, for the first time using Order~$N$ techniques, the
photonic band structure for photonic materials containing metal. Our
work represents a significant extention for Chan's Order~$N$ band
structure method. We expect to be able to handle a wide range of other
dispersive materials. Within the stated limits, the scheme exhibits
excellent numerical stability, in common with Chan's original scheme.
Our method
retains the advantageous order~$N$ scaling and is therefore capable of
treating complex structures with large unit cells.

\begin{figure}
\caption{Photonic band structure for a periodic series of infinite metallic
sheets. The filling fraction $f=0.1$, and 
 $w_{p}d/2\pi c = 1$. Calculated using the Order~$N$ method.}
\label{fig:1d}
\end{figure}

\begin{figure}
\caption{Photonic band structure for a periodic series of infinite metallic
sheets. The filling fraction $f=0.1$, and
 $w_{p}d/2\pi c = 1$. Calculated using the transfer matrix method.}
\label{fig:1d_tmm}
\end{figure}

\begin{figure}
\caption{The width of the gap below the lowest frequency band as a
function of the filling fraction filling fraction. For the one
dimensional system of metallic sheets.}
\label{fig:gap}
\end{figure}

\begin{figure}
\caption{The photonic band structure for a square lattice of metal
rods with a square cross-section. The filling fraction $f=0.01$, and
 $w_{p}d/2\pi c = 1$. Calculated using the Order~$N$ method.}
\label{fig:2d}
\end{figure}

\begin{figure}
\caption{The photonic band structure for a square lattice of metal
rods with a square cross-section. The filling fraction $f=0.01$, and
 $w_{p}d/2\pi c = 1$. Calculated using the transfer matrix method.}
\label{fig:2d_tmm}
\end{figure}

\begin{figure}
\caption{The photonic band structure for a square lattice of metal
rods with a square cross-section. The filling fraction $f=0.04$, and
 $w_{p}d/2\pi c = 1$. Calculated using the Order~$N$ method.}
\label{fig:2dH}
\end{figure}

\begin{figure}
\caption{The photonic band structure for a square lattice of metal
rods with a square cross-section. The filling fraction $f=0.04$, and
 $w_{p}d/2\pi c = 1$. Calculated using the transfer matrix method.}
\label{fig:2dH_tmm}
\end{figure}

\end{document}